\documentclass[12pt]{iopart}

\usepackage{iopams}
\usepackage{graphicx}

\begin{document}

\title[Non-Markovian dynamics of interacting qubit \ldots]
{Non-Markovian dynamics of interacting qubit pair coupled to two
independent bosonic baths}

\author{I. Sinayskiy}

\address{Quantum Research Group, School of Physics and National
Institute for Theoretical Physics, University of KwaZulu-Natal,
Durban, 4001, South Africa} \ead{ilsinay@gmail.com}

\author{E. Ferraro}
\address{MIUR and Dipartimento di Scienze Fisiche ed
Astronomiche dell'Universit\`{a} di Palermo, via Archirafi 36,
I-90123 Palermo, Italy} \ead{ferraro@fisica.unipa.it}

\author{A. Napoli}
\address{MIUR and Dipartimento di Scienze Fisiche ed
Astronomiche dell'Universit\`{a} di Palermo, via Archirafi 36,
I-90123 Palermo, Italy} \ead{napoli@fisica.unipa.it}

\author{A. Messina}

\address{MIUR and Dipartimento di Scienze Fisiche ed
Astronomiche dell'Universit\`{a} di Palermo, via Archirafi 36,
I-90123 Palermo, Italy} \ead{messina@fisica.unipa.it}

\author{F. Petruccione}

\address{Quantum Research Group, School of Physics and National
Institute for Theoretical Physics, University of KwaZulu-Natal,
Durban, 4001, South Africa} \ead{petruccione@ukzn.ac.za}

\begin{abstract}
The dynamics of two weakly interacting spins coupled to separate
bosonic baths is studied. An analytical solution in Born
approximation for arbitrary spectral density functions of the
bosonic environments is found. It is shown that in the
non-Markovian cases concurrence ``lives" longer or reaches greater
values.
\end{abstract}


\maketitle

\section{Introduction}

The implementation of more and more efficient nanodevices
exploitable in applicative contexts like quantum computers, often
requires a highly challenging miniaturization process aimed at
packing a huge number of point-like basic elements, whose
dynamics mimics indeed that of a qubit. Stimulated by such a
requirement, over the last few years theoretical schemes have
been investigated in the language of spin $\frac{1}{2}$ models
\cite{amico}. Apart from the simple dynamical behaviour of each
elementary constituent these Hamiltonian models do indeed capture
basic ingredients of several physical situations. In addition,
spin models allow for the description of the effective
interactions in a variety of different physical contexts ranging
from high energy to nuclear physics \cite{polyakov,belitsky}. In
condensed matter physics they capture several aspects of
high-temperature superconductors, quantum Hall systems, and heavy
fermions \cite{Prokofev,Privman,Kane}. We point out that
Hamiltonians for interacting spins can be realized artificially
in Josephson junctions arrays \cite{fazio} or with neutral atoms
loaded in optical lattices \cite{duan,jané,porras} or else with
electrons in quantum dots \cite{Schliemann}.

In this context, a subject deserving a particular interest is the
entanglement dynamics. In view of possible applications it is
important to understand the extent at which quantum coherences
may be protected against the unavoidable degradation of the
purity of the state, in particular in the presence of many-body
interactions.

In this paper we focus our attention on a spin model recently
introduced by Quiroga \cite{quiroga} and successively analyzed by
other authors \cite{sinaysky}-\cite{scala}. It consists of two
interacting spins $\frac{1}{2}$, each one coupled to a separate
bosonic bath \cite{quiroga,sinaysky}. Our aim is to study the
entanglement dynamics of the two spins in the non-Markovian
regime. Many authors have addressed the question of the dynamics
of the entanglement between qubits in the non-Markovian
environments. However, usually a system of non-interacting qubits
in contact with separate bosonic baths is considered. Either,
entanglement is introduced in the initial preparation
\cite{C1,C2} or created by the interaction of qubits with a
common environment \cite{cui}. The focus of this paper is to
study a system of directly interacting qubits. This is a typical
situation in solid-state systems. For example, double quantum
dots can be modeled as coupled qubit systems in contact with
separate bosonic baths. For the demonstration of the dynamical
properties of the system, in this paper we will consider Lorentz
spectral density and Ohmic spectral density with Lorentz-Drude
cut-off. For a different model it has been shown that
entanglement of qubits can occur in super-Ohmic environments even
at non-vanishinig temperature \cite{L1,L2}.

The paper is structured as follows. In Section 2 we describe in
detail the model. In Section 3 we present the analytical solution
of the non-Markovian master equation for the reduced system
constituted by the two interacting spins in the zero temperature
limit. In Section 4 we analyze the entanglement dynamics of the
two spins assuming for the environment a Lorentz spectral density
and an Ohmic spectral density with a Lorentz-Drude cut-off
function. Finally, conclusive remarks are given in Section 5.

\section{The model}
Our analysis is focused on the dynamics of a composite system
coupled to bosonic environments. Parts of the dynamical system are
weakly interacting. The total Hamiltonian can be written as:
\begin{equation}\label{eq:HTotal}
H=H_S+\lambda^2 H_I+H_B+\lambda H_{SB},
\end{equation}
where $(H_S+\lambda^2 H_I)$ is the Hamiltonian describing the
dynamics of the composite system, $H_S$ is the Hamiltonian of the
free components of the system, $H_I$ is the Hamiltonian of
interaction between the parts of the system. The operator $H_B$
describes bosonic environments, the Hamiltonian $H_{SB}$ denotes
the Hamiltonian of the interaction between system and environment.
The parameter $\lambda$ is a dimensionless expansion parameter.
The non-Markovian dynamics of the reduced system will be
described by a Master Equation containing the terms not higher
than the square of the expansion parameter $\lambda$.

The second-order time-convolutionless form of the Master equation
is given by \cite{petruccionebook}:
\begin{equation}\label{eq:tcl2gen}
\frac{d}{dt}\rho^I_S(t)=-\lambda^2\int_0^td\tau\mathrm{tr_B}[H_{SB}(t),[H_{SB}(\tau),\rho^I_S(t)\otimes\rho_B]],
\end{equation}
where $H_{SB}(t)$ denotes the Hamiltonian $H_{SB}$ and
$\rho^I_S(t)$ denotes the density matrix of the reduced system in
the interaction picture by the Hamiltonian $(H_S+\lambda^2
H_I+H_B)$. The density matrix $\rho_B=e^{-\beta
H_B}/\mathrm{tr}[e^{-\beta H_B}]$ describes the state of the
environment.

The present general approach is applied to a system consisting of
a pair of weakly interacting spins, each one coupled to a bosonic
bath. The total Hamiltonian is given by Eq. (\ref{eq:HTotal}).
The Hamiltonian of the two free spins characterized by the same
energy $\epsilon$ reads
\begin{equation}
H_S=\frac{\epsilon}{2}\sigma^z_1+\frac{\epsilon}{2}\sigma^z_2.
\end{equation}
As usual $\sigma_i^z$ and $\sigma_i^{\pm}$ are the Pauli
operators describing the $i-$th spin $(i=1,2)$. The Hamiltonian
of the weakly interacting spins is given by
\begin{equation}\label{eq:HS}
\lambda^2 H_I
=K\left(\sigma^+_1\sigma^-_2+\sigma^-_1\sigma^+_2\right),
\end{equation}
where $K$ is a constant defining the strength of the spin-spin
interaction. The Hamiltonian of the bosonic baths characterized
by the annihilation and creation operators $b_{ni}$ and
$b_{ni}^{\dagger}$ $(i=1,2)$ reads
\begin{equation}
H_B=\sum_n\omega_{n,1}b^\dag_{n,1}b_{n,1}+\sum_m
\omega_{m,2}b^\dag_{m,2}b_{m,2}.
\end{equation}
The coupling of each spin to the separate bosonic baths is
described by
\begin{equation}
H_{SB}
=\sigma^+_1\sum_ng_{n,1}b_{n,1}+\sigma^+_2\sum_mg_{m,2}b_{m,2}+\mathrm{h.c.},
\end{equation}
where $g_{n,1}$ and $g_{m,2}$ denote the coupling between the spin
and its corresponding bosonic baths. In this paper units are
chosen such that $k_B=\hbar=1$. The Hamiltonian $\lambda H_{SB}$
in the interaction picture defined by the Hamiltonian
$(H_S+\lambda^2 H_I+H_B)$ is given by
\begin{equation}\label{hd}
\lambda
H_{SB}(t)=\sigma^+_1\sum_ng_{n,1}b_{n,1}e^{i\left(\epsilon-\omega_{n,1}\right)t}+\sigma^+_2\sum_ng_{n,2}b_{n,2}e^{i\left(\epsilon-\omega_{n,2}\right)t}+
\mathrm{h.c.}.
\end{equation}
In the above expression we neglect terms proportional to the cube
of $\lambda$ and higher. By direct calculation we show that
\begin{equation}\label{eq:METCL2}
-\lambda^2\int_0^td\tau\mathrm{tr_B}[H_{SB}(t),[H_{SB}(\tau),\rho^I_S(t)\otimes\rho_B]]=\sum_{j=1}^2\mathcal{L}^{(Dj)}(t)\rho^I_S(t),
\end{equation}
where $\mathcal{L}^{(Dj)}(t)$ is the Liouville superoperator
defined by
\begin{eqnarray}
\mathcal{L}^{(Dj)}\rho^I_S(t)&=&B^{(j)}(t)\left[\sigma^-_j\rho_S(t),\sigma^+_j\right]+\bar{B}^{(j)}(t)\left[\sigma^-_j,\rho_S(t)\sigma^+_j\right]\\&\nonumber
&
+\bar{A}^{(j)}(t)\left[\sigma^+_j\rho_S(t),\sigma^-_j\right]+A^{(j)}(t)\left[\sigma^+_j,\rho_S(t)\sigma^-_j\right].
\end{eqnarray}
The quantities $A^{(j)}(t)$ and $B^{(j)}(t)$ appearing in the
previous expression are the so-called correlation functions,
whose explicit form is given by
\begin{eqnarray}
A^{(j)}(t)&=&\int_0^td\tau\sum_n|g_{n,j}|^2\langle
b^\dag_{n,j}b_{n,j}\rangle_{Bj}e^{i\left(\epsilon-\omega_{n,j}\right)(t-\tau)}\\\nonumber
&=&i\sum_n|g_{n,j}|^2\langle
b^\dag_{n,j}b_{n,j}\rangle_{Bj}\frac{1-e^{i\left(\epsilon-\omega_{n,j}\right)t}}{\epsilon-\omega_{n,j}},
\end{eqnarray}
\begin{eqnarray}\label{eq:CFB}
B^{(j)}(t)&=&\int_0^td\tau\sum_n|g_{n,j}|^2\langle
b_{n,j}b^\dag_{n,j}\rangle_{Bj}e^{i\left(\epsilon-\omega_{n,j}\right)(t-\tau)}\\\nonumber
&=&i\sum_n|g_{n,j}|^2\langle
b_{n,j}b^\dag_{n,j}\rangle_{Bj}\frac{1-e^{i\left(\epsilon-\omega_{n,j}\right)t}}{\epsilon-\omega_{n,j}},
\end{eqnarray}
where $\langle O\rangle_{Bj}\equiv
\mathrm{tr}_{B_j}\{O\rho_{Bj}\}$, $\bar{A}^{(j)}(t)$ and
$\bar{B}^{(j)}(t)$ being the complex conjugate of $A^{(j)}(t)$
and $B^{(j)}(t)$, respectively. To obtain expression
(\ref{eq:METCL2}) we used the fact that the bosonic environments
assumed in this article are uncorrelated with each other and
$\langle b_{n,j}b^\dag_{n,j}\rangle_{Bj},\langle
b^\dag_{n,j}b_{n,j}\rangle_{Bj}$ are the only non-zero
second-order correlations in the bath, all the other vanish.

Transforming back to the Schr\"odinger picture we obtain the
following Master Equation
\begin{equation}\label{eq:ME1}
\frac{d}{dt}\rho_S(t)=-i[\frac{\epsilon}{2}\sigma^z_1+\frac{\epsilon}{2}\sigma^z_2+K\left(\sigma^+_1\sigma^-_2+\sigma^-_1\sigma^+_2\right),\rho_S(t)]+\sum_{j=1}^2\mathcal{L}^{(Dj)}(t)\rho_S(t).
\end{equation}
It is easy to see that the superoperator $\mathcal{L}_0$ defined
as
\begin{equation}
\mathcal{L}_0\rho_S(t)=-i[\frac{\epsilon}{2}\sigma^z_1+\frac{\epsilon}{2}\sigma^z_2,\rho_S(t)]
\end{equation}
commutes with the superoperator $\mathcal{L}_{ME}(t)$ given by
\begin{equation}
\mathcal{L}_{ME}(t)\rho_S(t)=-i[K\left(\sigma^+_1\sigma^-_2+\sigma^-_1\sigma^+_2\right),\rho_S(t)]+\sum_{j=1}^2\mathcal{L}^{(Dj)}(t)\rho_S(t),
\end{equation}
and can be neglected as it is irrelevant for the dynamics of the
expectation values defined by the density matrix $\rho_S(t)$. So,
the final form of the Master Equation which is going to be
studied in this article reads
\begin{equation}\label{eq:MEMain}
\frac{d}{dt}\rho_S(t)=-i[K\left(\sigma^+_1\sigma^-_2+\sigma^-_1\sigma^+_2\right),\rho_S(t)]+\sum_{j=1}^2\mathcal{L}^{(Dj)}(t)\rho_S(t).
\end{equation}

\section{Exact solution of the Master equation}
In order to solve the Master equation (\ref{eq:MEMain}), it is
 useful to separate the equations of motion for the diagonal elements of the density operator
$\rho_S(t)$ from those relative to the off-diagonal elements. We
have indeed proved that the diagonal and two non-diagonal
elements of $\rho_S(t)$ have to satisfy the following system of
the equations
\begin{equation}
\frac{d}{dt}\left(\begin{array}{c}
\rho_S^{11}(t)\\
\rho_S^{22}(t)\\
\rho_S^{33}(t)\\
\rho_S^{44}(t)\\
\rho_S^{23}(t)\\
\rho_S^{32}(t)\end{array}\right)=\Lambda_6(t)\left(\begin{array}{c}
\rho_S^{11}(t)\\
\rho_S^{22}(t)\\
\rho_S^{33}(t)\\
\rho_S^{44}(t)\\
\rho_S^{23}(t)\\
\rho_S^{32}(t)\end{array}\right),
\end{equation}
where
\begin{equation}
\hspace{-1cm}\Lambda_6(t)=\left(\begin{array}{cccccc}
-\beta_1-\beta_2 & \alpha_2 & \alpha_1 & 0 & 0 & 0\\
\beta_2 & -\alpha_2-\beta_1  & 0 & \alpha_1 & iK & -iK\\
\beta_1 & 0 & -\alpha_1-\beta_2 & \alpha_2 & -iK & iK\\
0 & \beta_1 & \beta_2 & -\alpha_1-\alpha_2 & 0 & 0\\
0 & iK & -iK & 0 & \xi & 0\\
0 & -iK & iK & 0 & 0 & \bar{\xi}
\end{array}\right)
\end{equation}
and
\begin{eqnarray}
\fl
\hspace{2cm}\alpha_j=A^{(j)}(t)+\bar{A}^{(j)}(t),\quad\beta_j=B^{(j)}(t)+\bar{B}^{(j)}(t),\\\nonumber
\xi=-A^{(1)}(t)-\bar{A}^{(2)}(t)-B^{(1)}(t)-\bar{B}^{(2)}(t).
\end{eqnarray}
In what follows we will consider the case in which the two bosonic
baths are both prepared in a thermal state with $T=0$. This
assumption in turn implies that the correlation functions reduce to
\begin{equation}
A^{(j)}(t)\equiv0,\quad B^{(j)}(t)\equiv
B(t)=i\sum_n|g_{n}|^2\frac{1-e^{i\left(\epsilon-\omega_{n}\right)t}}{\epsilon-\omega_{n}}.
\end{equation}
Under these hypotheses it is possible to rewrite $\Lambda_6(t)$ in
the following way, $\Lambda_6(t)=(B(t)+\bar{B}(t))L_1+iKL_2$,
where $L_1$ and $L_2$ are $6\times 6$ commuting matrices. Thus,
the solution of the previous system of differential equations can
be written as
\begin{equation}
\left(\begin{array}{c}
\rho_S^{11}(t)\\
\rho_S^{22}(t)\\
\rho_S^{33}(t)\\
\rho_S^{44}(t)\\
\rho_S^{23}(t)\\
\rho_S^{32}(t)\end{array}\right)=U_6(t)\left(\begin{array}{c}
\rho_S^{11}(0)\\
\rho_S^{22}(0)\\
\rho_S^{33}(0)\\
\rho_S^{44}(0)\\
\rho_S^{23}(0)\\
\rho_S^{32}(0)\end{array}\right),
\end{equation}
where
\begin{equation}
U_6(t)=\mathrm{T}e^{\int_0^t d\tau
\Lambda_6(\tau)}=e^{G(t)L_1}e^{(iKt)L_2}
\end{equation}
and the symbol $\mathrm{T}$ denotes the standard time-ordering in
the exponent. The function $G(t)$ appearing in the expression for
the matrix $U^{(6)}(t)$ is defined as
\begin{equation}
G(t)=\Phi(t)+\bar{\Phi}(t),
\end{equation}
with
\begin{equation}
\Phi(t)=\int_0^td\tau
B(\tau)=\sum_n|g_{n}|^2\frac{1-e^{i\left(\epsilon-\omega_{n}\right)t}+i\left(\epsilon-\omega_{n}\right)t}{\left(\epsilon-\omega_{n}\right)^2}.
\end{equation}
The time dependence of the off-diagonal element $\rho_S^{14}(t)$
is trivial, namely
$\rho_S^{14}(t)=\exp{(-2\Phi(t))}\rho_S^{14}(0)$. For the other
off-diagonal elements we get the following system of equations:
\begin{equation}
\frac{d}{dt}\left(\begin{array}{c}
\rho_S^{12}(t)\\
\rho_S^{13}(t)\\
\rho_S^{24}(t)\\
\rho_S^{34}(t)\end{array}\right)=\Lambda_4(t)\left(\begin{array}{c}
\rho_S^{12}(t)\\
\rho_S^{13}(t)\\
\rho_S^{24}(t)\\
\rho_S^{34}(t)\end{array}\right),
\end{equation}
where
\begin{equation}
\hspace{-1cm}\Lambda_4(t)=\left(\begin{array}{cccc}
-\beta-B(t) & iK & 0 & 0 \\
iK & \beta-B(t) & 0 & 0 \\
0 & \beta & -B(t) & -iK \\
\beta & 0 & -iK & -B(t) \\
\end{array}\right).
\end{equation}
One can check that the solution for the above equation has the
following form:
\begin{equation}
\left(\begin{array}{c}
\rho_S^{12}(t)\\
\rho_S^{13}(t)\end{array}\right)=e^{-G(t)-\Phi(t)}e^{iKt\sigma_x}\left(\begin{array}{c}
\rho_S^{12}(0)\\
\rho_S^{13}(0)\end{array}\right)
\end{equation}
and
\begin{equation}
\left(\begin{array}{c}
\rho_S^{24}(t)\\
\rho_S^{34}(t)\end{array}\right)=U_2(t)\left(\begin{array}{c}
\rho_S^{24}(0)\\
\rho_S^{34}(0)\end{array}\right)+U_2(t)\int_0^td\tau
U_2^{-1}(\tau) \left(\begin{array}{c}
\rho_S^{13}(\tau)\\
\rho_S^{12}(\tau)\end{array}\right),
\end{equation}
where the operator $U_2(t)$ is defined by
\begin{equation}
U_2(t)=e^{-\Phi(t)}e^{-iKt\sigma_x}.
\end{equation}

At this point we are in the position to explicitly write the
density matrix of the two coupled spins at a generic time $t$
starting from an arbitrary initial condition. For simplicity, we
report on such a solution in the Appendix. In what follows,
instead, we focus on the cases in which the initial state of the
pair of coupled spins is the Bell state
$|\Psi_-\rangle=\frac{1}{\sqrt{2}}(|10\rangle-|01\rangle)$ or the
factorized state $|\Psi_0\rangle=|10\rangle$. Exploiting the
results presented in the Appendix it is possible to demonstrate
that the state of the reduced system at a generic time instant
$t$ can be written in the simple form
\begin{equation}\label{rho}
\rho_{S(\mathrm{Bell})}(t)=e^{-G(t)}|\Psi_-\rangle\langle\Psi_-|+\left(1-e^{-G(t)}\right)|00\rangle\langle00|
\end{equation}
and
\begin{equation}\label{rho2}
\rho_{S(0)}(t)=e^{-G(t)}|\Psi(t)\rangle\langle\Psi(t)|+\left(1-e^{-G(t)}\right)|00\rangle\langle00|,
\end{equation}
where
\begin{equation}
|\Psi(t)\rangle=\cos(Kt)|10\rangle-i\sin(Kt)|01\rangle.
\end{equation}

Another point which we would like to mention here is the
connection between the non-Markovian Master Equation
(\ref{eq:MEMain}) and the Markovian one. The Markovian limit of
the Master Equation (\ref{eq:MEMain}) can be constructed by taking
the limit $t\rightarrow\infty$ in the set of correlation functions
$A^{(j)}(t)$ and $B^{(j)}(t)$. The solution of the corresponding
Markovian Master Equation for the system at hand can be
constructed from non-Markovian ones by replacing functions
$\Phi(t)$ and $G(t)$ with the corresponding Markovian ones
\begin{equation}
\Phi(t)\Rightarrow\Phi^M(t)=tB^M,
\end{equation}
where
\begin{equation}
B^M=\lim_{t\rightarrow\infty}B(t).
\end{equation}
In particular, for the function $G(t)$ we have
\begin{equation}\label{eq:GM}
G(t)\Rightarrow G^M(t)=t(B^M+\bar{B}^M)=t 2\pi J(\omega_0),
\end{equation}
where $J(\omega_0)$ is the bath spectral density and
$\omega_0=\frac{\epsilon}{2}$.

\section{Entanglement dynamics}
As emphasized before, the solution we have found has been obtained
without specifying the spectral properties of the bath. The
density matrix $\rho_S(t)$ describing the pair of the coupled
spins, however, depends on the bath spectral density through the
function $G(t)$.

\begin{figure}
\begin{center}
\includegraphics[width=0.9\linewidth]{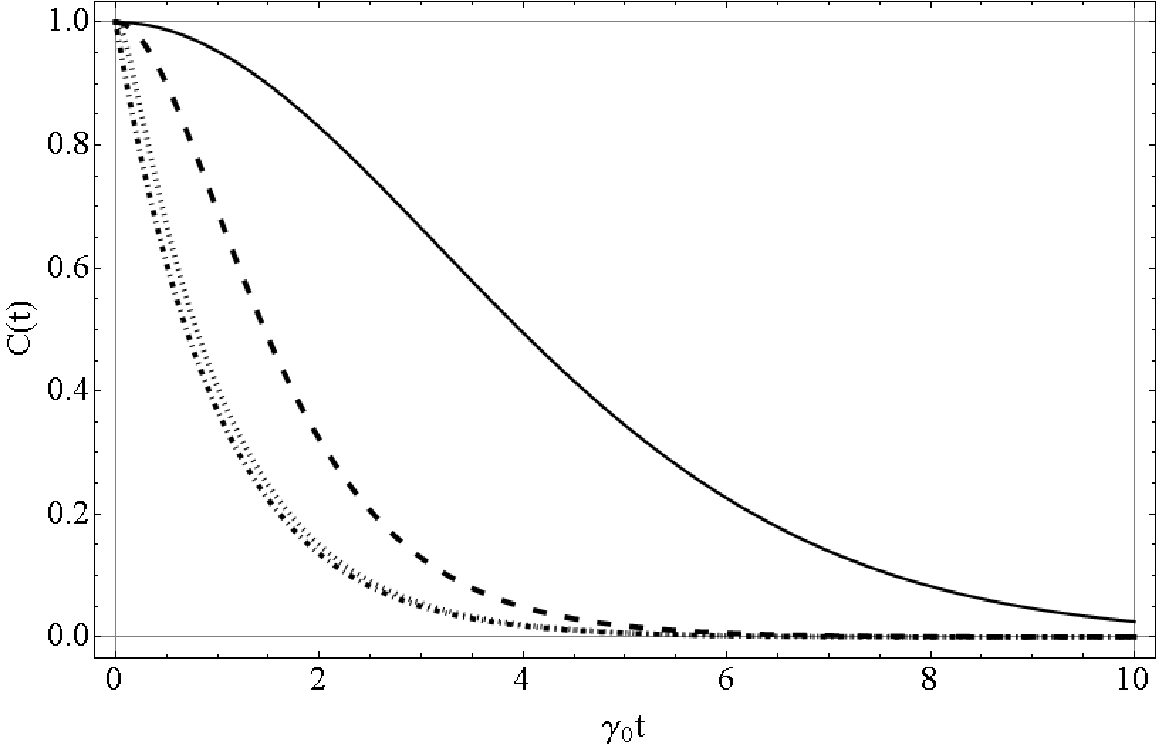}
\caption{The concurrence $C(t)$ for a Lorentz bath distribution
for different values of the ratio $\gamma/\gamma_0$
($\gamma/\gamma_0=0.1$ (solid line), $\gamma/\gamma_0=1$ (dashed
line), $\gamma/\gamma_0=10$ (dotted line), Markovian case
(dot-and-dash line)). The initial state is the Bell state
(\ref{bell0}).}\label{fig1}
\end{center}
\end{figure}

In this section, exploiting our results, we will analyze some
dynamical properties of the central system for different spectral
distributions of the environment. In particular, we will examine
how the entanglement evolution is affected by the choice of the
reservoir spectral density. Let us start by considering as a first
case the Lorentzian distribution
\begin{equation}\label{lorentz}
J(\omega)=\frac{\gamma_0}{2\pi}\frac{\gamma^2}{(\omega-\frac{\epsilon}{2})^2+\gamma^2},
\end{equation}
where $\gamma$ and $\gamma_0$ are the reservoir and the system
decay rate respectively. This choice in turn implies that the
correlation function $B(t)$, as given in the previous section, is
\begin{equation}
B(t)=\frac{\gamma_0}{2}(1-e^{-\gamma t})
\end{equation}
and consequently
\begin{equation}
G(t)=\gamma_0t+\frac{\gamma_0}{\gamma}(e^{-\gamma t}-1).
\end{equation}
We have already demonstrated that starting from the Bell state
\begin{equation}\label{bell0}
|\psi_-\rangle=\frac{1}{\sqrt{2}}(|10\rangle-|01\rangle)
\end{equation}
at a generic time $t$ the density operator describing our system
can be written as in Eq.(\ref{rho}). It is interesting to analyze
how the interaction of the two coupled spins with the
environments modifies the entanglement initially present in the
system. To this end we consider the time behavior of the
concurrence \cite{wootters} of the two spins. Using Eq.(\ref{rho})
it is easy to demonstrate that in correspondence to any
environmental spectral density, the concurrence is given by
\begin{equation}\label{conc}
C(t)=e^{-G(t)}.
\end{equation}
Thus, when $J(\omega)$ assumes the form (\ref{lorentz}) we have
\begin{equation}
C(t)=\exp{\left(-(\gamma_0t+\frac{\gamma_0}{\gamma}(e^{-\gamma
t}-1))\right)}.
\end{equation}
In Figure \ref{fig1} we plot $C(t)$ against $\gamma_0 t$ for
different values of the ratio $\gamma/\gamma_0$ in the
non-Markovian case. For comparison with the Markovian limit
(\ref{eq:GM}) we include also the Markovian case ($G^M(t)=\gamma_0
t$).
As expected, in the presence of the two baths the concurrence
function, starting from its maximum value, decreases as time
elapses. However, in the non-Markovian regime, corresponding to
$\gamma/\gamma_0<2$, the entanglement in the two spins persists
for a longer time with respect to the Markovian case.

Suppose now that the two environments are characterized by an
Ohmic spectral density with a Lorentz-Drude cut-off function
\cite{cui,MPIPM}
\begin{equation}\label{LD}
J(\omega)=\frac{2\omega}{\pi}\frac{\omega_c^2}{\omega_c^2+\omega^2},
\end{equation}
where $\omega$ is the frequency of the bath and $\omega_c$ is the
cut-off frequency. Under this hypothesis, putting
$\omega_0=\epsilon/2$, the correlation function becomes
\begin{equation}
B(t)=-i\frac{2\omega_c^2}{\omega_c-i\omega_0}(1-e^{-(\omega_c-i\omega_0)t})
\end{equation}
and thus
\begin{eqnarray}
G(t)&=4\frac{\omega_c^2\omega_0}{\omega_c^2+\omega_0^2}t+
4\frac{\omega_c^2}{(\omega_c^2+\omega_0^2)^2}(\omega_c^2-\omega_0^2)e^{-\omega_ct}\sin(\omega_0t)+\nonumber\\
&+8\frac{\omega_c^3\omega_0}{(\omega_c^2+\omega_0^2)^2}e^{-\omega_ct}\cos(\omega_0t)-
8\frac{\omega_c^3\omega_0}{(\omega_c^2+\omega_0^2)^2}.
\end{eqnarray}
The corresponding Markovian function reads
\begin{equation}
G^M(t)=2\pi J(\omega_0)t
=4\frac{\omega_c^2\omega_0}{\omega_c^2+\omega_0^2}t.
\end{equation}

\begin{figure}
\begin{center}
\includegraphics[width=0.85\linewidth]{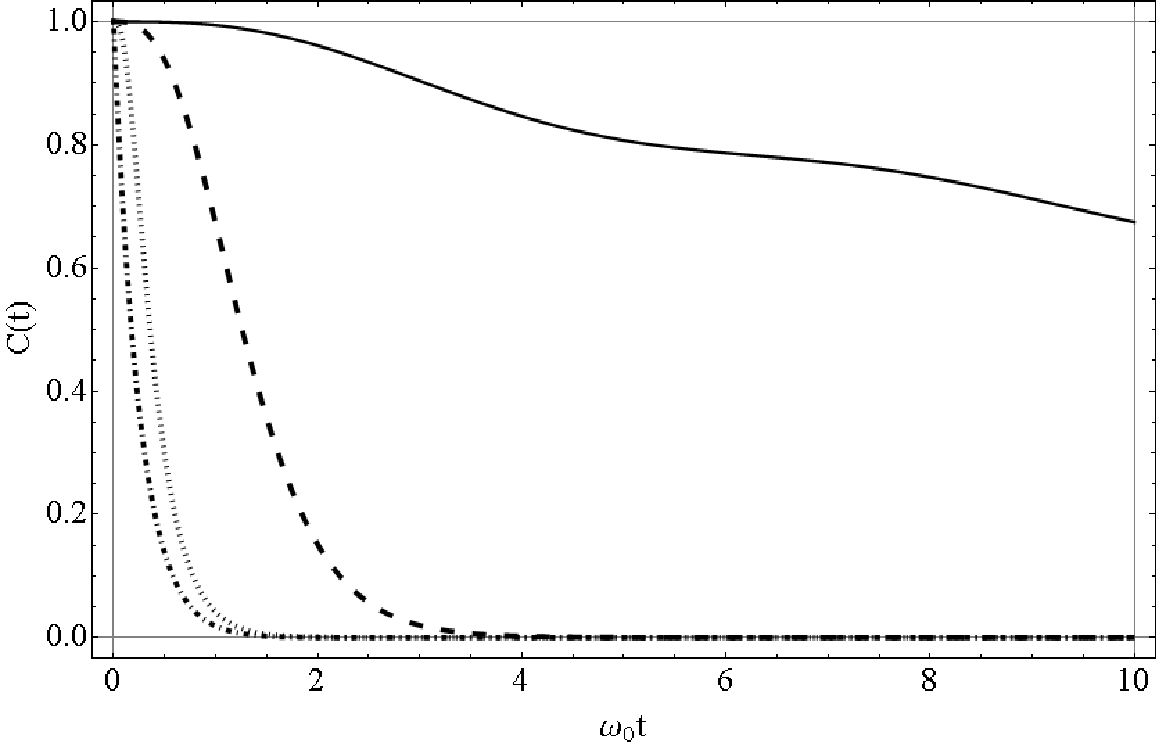}
\caption{The concurrence $C(t)$ for a Lorentz-Drude bath
distribution for different values of the ratio $\omega_c/\omega_0$
($\omega_c/\omega_0=0.1$ (solid line), $\omega_c/\omega_0=1$
(dashed line), $\omega_c/\omega_0=10$ (dotted line), Markovian
case (dot-and-dash line)). The initial state is the Bell state
(\ref{bell0}).}\label{fig2}
\end{center}
\end{figure}

Using Eq.(\ref{conc}) it is possible to analyze the evolution of
the degree of entanglement of the two spins starting from the Bell
state (\ref{bell0}). The results we have obtained are reported in
Figure \ref{fig2} for different values of the ratio
$\omega_c/\omega_0$.
Comparing the four plots, we may observe that when the spectrum of
the reservoir does not completely overlap with the frequency of
the system, that is $\omega_c\ll\omega_0$, the concurrence
decreases to zero more slowly than in the opposite case,
$\omega_c\gg\omega_0$. The results we have obtained, reported in
Figures \ref{fig1} and \ref{fig2}, indicate that when the baths
are characterized by Ohmic spectral densities with a
Lorentz-Drude cut-off function, as given in Eq.(\ref{LD}), the
entanglement initially present in the two spins can be preserved
for a longer time with respect to the case of a Lorentz bath, at
least for some values of the ratio $\omega_c/\omega_0$.

\begin{figure}
\begin{center}
\includegraphics[width=0.8\linewidth]{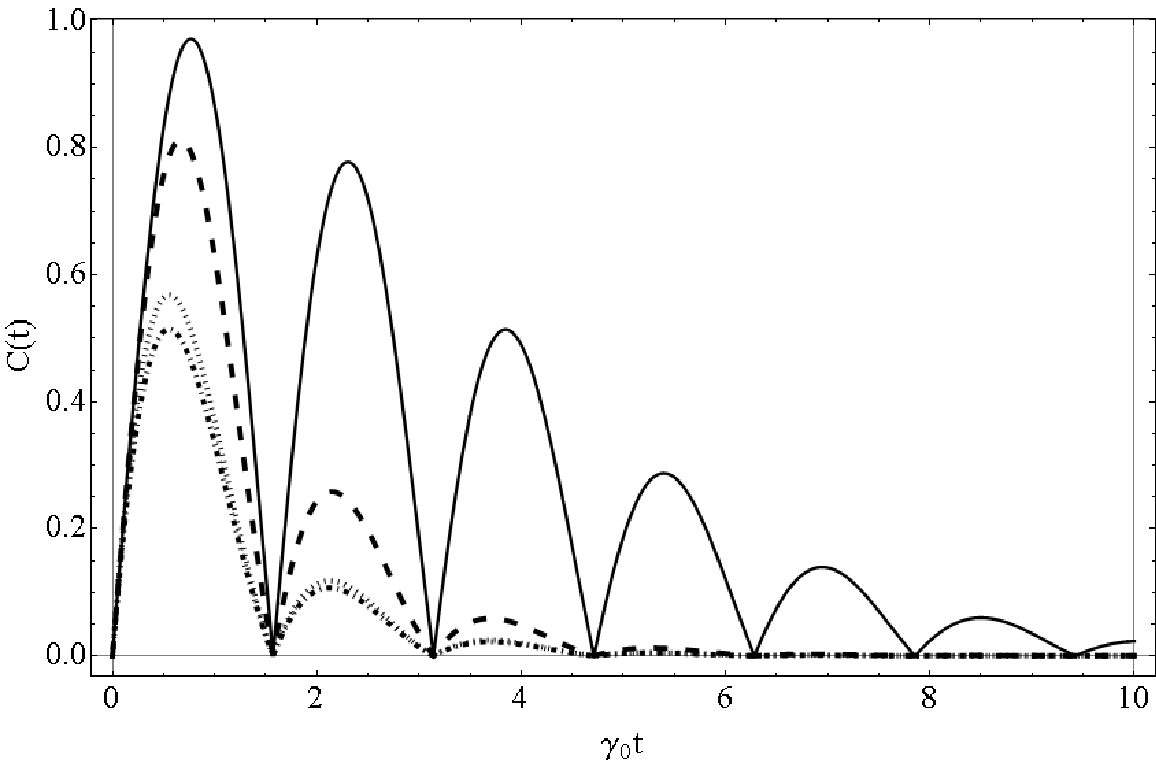}
\caption{The concurrence $C(t)$ for a Lorentz bath distribution
for different values of the ratio $\gamma/\gamma_0$
($\gamma/\gamma_0=0.1$ (solid line), $\gamma/\gamma_0=1$ (dashed
line), $\gamma/\gamma_0=10$ (dotted line), Markovian case
(dot-and-dash line)). The initial state is
$|1,0\rangle$.}\label{fig3}
\end{center}
\end{figure}

\begin{figure}
\begin{center}
\includegraphics[width=0.9\linewidth]{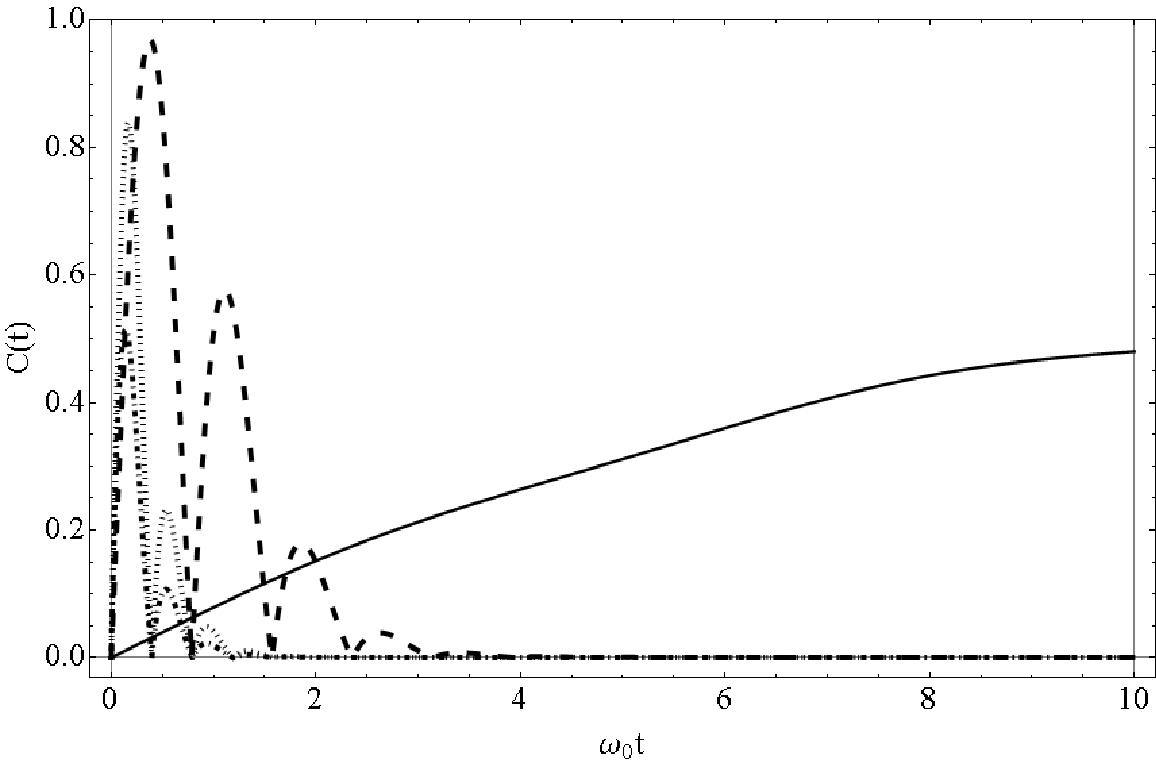}
\caption{The concurrence $C(t)$ for a Lorentz-Drude bath
distribution for different values of the ratio $\omega_c/\omega_0$
($\omega_c/\omega_0=0.1$ (solid line), $\omega_c/\omega_0=1$
(dashed line), $\omega_c/\omega_0=10$ (dotted line), Markovian
case (dot-and-dash line)). The initial state is
$|1,0\rangle$.}\label{fig4}
\end{center}
\end{figure}

Following the analysis developed in this section it is also
interesting to examine the behavior of the system starting from a
factorized initial condition instead of an entangled one. In what
follows, in particular, we suppose that at $t=0$ the two spins are
in the separable state $|1,0\rangle$ and we study the time
behaviour of the concurrence. We find that in this case
\begin{equation}\label{conc2}
C(t)=e^{-G(t)}|\sin(2Kt)|.
\end{equation}
The interaction between the two spins, as expressed by the
effective Hamiltonian (\ref{eq:HS}), enables the generation of
entanglement starting from the factorized initial condition given
before. On the other hand, in view of the fact that the two spins
are coupled to two different baths, the quantum correlations that
are established in the pair of spins will be destroyed. In the
non-Markovian regime, however, we expect that the entanglement
will be preserved for a longer time with respect to the Markovian
one. This is confirmed by the time behaviour of the concurrence
function of the two spins for the Lorentzian spectral density of
the baths (Figure \ref{fig3}) and for the Ohmic spectral density
of the baths (Figure \ref{fig4}). Looking at these figures we
also observe that the degree of entanglement that we can realize
in the system starting from the state $|1,0\rangle$ depends on
the ratio $\gamma/\gamma_0$ or $\omega_c/\omega_0$. In
particular, for the Lorentz spectral density, Figure \ref{fig3},
the maximum value of the concurrence function is reached in the
highly non-Markovian case, that is, $\gamma/\gamma_0=0.1$. For
the Ohmic spectral density, Figure \ref{fig4}, the highly
non-Markovian case ($\omega_c/\omega_0=0.1$) corresponds to the
presence of the quantum correlation in the system for the longest
time.

\begin{figure}
\begin{center}
\includegraphics[width=0.9\linewidth]{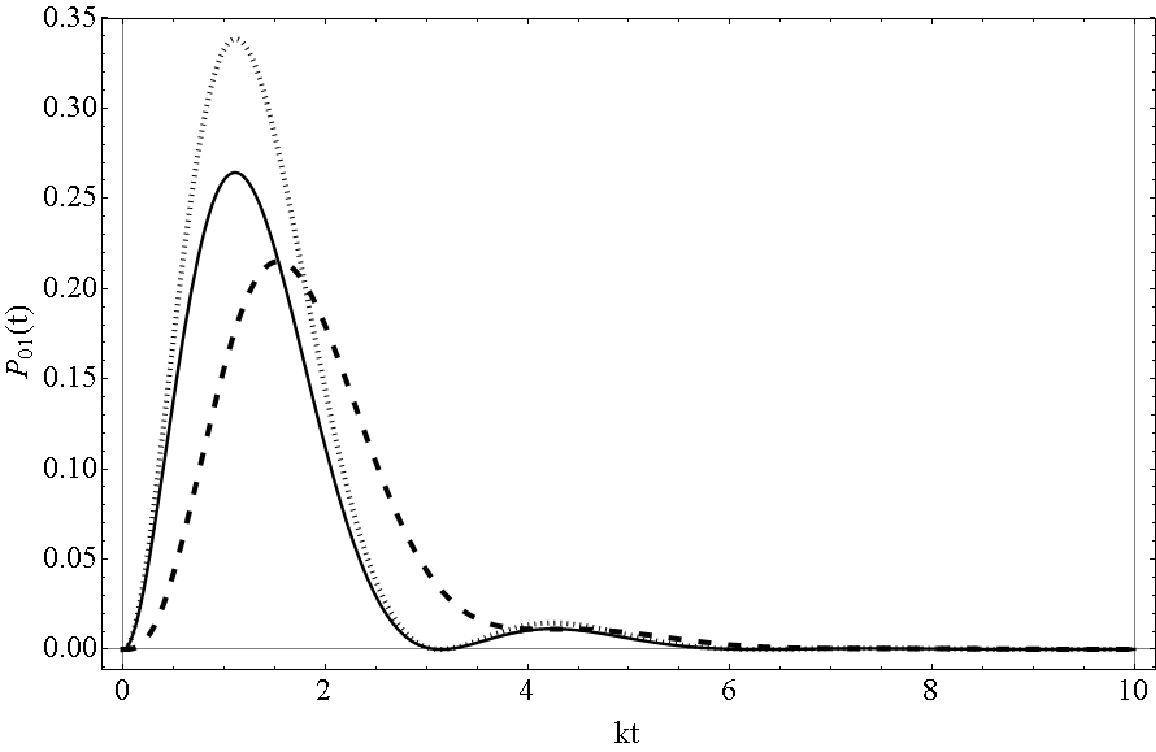}
\caption{Dynamics of the probability to find the system in the
state $|0,1\rangle$  (Markovian regime (solid line),
post-Markovian regime (dashed line), non-Markovian regime (dotted
line) for a Lorentz bath distribution with $\gamma/\gamma_0=4$).
The initial state is the separable state
$|1,0\rangle$.}\label{fig5}
\end{center}
\end{figure}

Before concluding we wish to compare our results with the ones
obtained in the Markovian \cite{sinaysky,scala} and post-Markovian
\cite{lidar,SP} regimes relatively to the same physical system. In
order to do this, we concentrate our attention on the temporal
behavior of the probability $P_{01}(t)$ of finding the qubit pair
in the state $|0,1\rangle$ supposing that at time $t=0$ the system
is prepared in the state $|1,0\rangle$. In Figure \ref{fig5},
where we show $P_{01}(t)$ in the three different regimes, time is
scaled in units of the strength $K$ of the spin-spin interaction.
As shown, when we are in the non-Markovian regime, $P_{01}(t)$
reaches a maximum value that is greater than the one
characterizing the Markovian and post-Markovian cases. Moreover,
as expected in view of the presence of the two baths, in all the
regimes the probability $P_{01}(t)$ decays toward zero after
reaching its maximum value.

\section{Conclusions}

In this paper we have analyzed the non-Markovian dynamics of a
pair of weakly interacting spins coupled to two separate bosonic
baths. After deriving the second-order master equation, that is
local in time, we have given an exact solved with the assumption
that the two bosonic environments are both prepared in a thermal
state with $T=0$. It is important to emphasize that our solution
is valid whatever the initial conditions of the system or the
spectral properties of the two baths may be. From the solution of
the non-Markovian Master Equation obtained we construct a solution
of the corresponding Master Equation in the Markovian limit.
Starting from the knowledge of the solution of master equation we
have studied the temporal behaviour of the entanglement
established in the pair of interacting spins for different
spectral densities. The results show that in the non-Markovian
case the concurrence, that is a measure of entanglement, of the
system of two spins ``lives" longer or reaches greater values
with respect to the Markovian regime. We wish to stress that the
results presented in the present paper are not directly connected
to the so-called ``entanglement sudden death" \cite{YE} because
the concurrence does not vanish for a certain finite instant of
time and has ``infinite" tails (\ref{conc}), (\ref{conc2}). Our
results motivate further studies on stronger coupling constants
and non-zero temperatures.

\section*{Acknowledgements}
This work is based upon research supported by the South African
Research Chair Initiative of the Department of Science and
Technology and National Research Foundation. AM (AN) acknowledges
partial support by MIUR project II04C0E3F3 (II04C1AF4E)
\textit{Collaborazioni Interuniversitarie ed Internazionali
tipologia C}.

\section*{Appendix}
\appendix
\setcounter{section}{1}

The full solution for the density matrix of the pair of spins for
arbitrary initial conditions reads:

\begin{eqnarray}
\fl \rho_S^{11}(t) = e^{-2G(t)}\rho_S^{11}(0),
\end{eqnarray}

\begin{eqnarray}
\fl \rho_S^{22}(t) =
e^{-G(t)}(1-e^{-G(t)})\rho_S^{11}(0)+e^{-G(t)}\cos^2(Kt)\rho_S^{22}(0)\\\nonumber+e^{-G(t)}\sin^2(Kt)\rho_S^{33}(0)-e^{-G(t)}\sin(2Kt)\mathrm{Im}(\rho_S^{23}(0)),
\end{eqnarray}

\begin{eqnarray}
\fl \rho_S^{33}(t) =
e^{-G(t)}(1-e^{-G(t)})\rho_S^{11}(0)+e^{-G(t)}\sin^2(Kt)\rho_S^{22}(0)\\\nonumber+e^{-G(t)}\cos^2(Kt)\rho_S^{33}(0)+e^{-G(t)}\sin(2Kt)\mathrm{Im}(\rho_S^{23}(0)),
\end{eqnarray}

\begin{eqnarray}
\fl \rho_S^{44}(t) = 1-\rho_S^{11}(t)-\rho_S^{22}(t)-\rho_S^{33}(t),
\end{eqnarray}

\begin{eqnarray}
\fl \rho_S^{23}(t) =
e^{-G(t)}\cos^2(Kt)\rho_S^{23}(0)+e^{-G(t)}\sin^2(Kt)\rho_S^{32}(0)\\\nonumber+\frac{i}{2}e^{-G(t)}\sin(2Kt)(\rho_S^{22}(0)-\rho_S^{33}(0)),
\end{eqnarray}

\begin{eqnarray}
\fl \rho_S^{14}(t) = e^{-2\Phi(t)}\rho_S^{14}(0),
\end{eqnarray}

\begin{eqnarray}
\fl \rho_S^{12}(t)
=e^{-G(t)-\Phi(t)}\cos(Kt)\rho_S^{12}(0)+ie^{-G(t)-\Phi(t)}\sin(Kt)\rho_S^{13}(0),
\end{eqnarray}

\begin{eqnarray}
\fl \rho_S^{13}(t)
=e^{-G(t)-\Phi(t)}\cos(Kt)\rho_S^{13}(0)+ie^{-G(t)-\Phi(t)}\sin(Kt)\rho_S^{12}(0),
\end{eqnarray}

\begin{eqnarray}
\fl \rho_S^{24}(t)
=e^{-\Phi(t)}\cos(Kt)\rho_S^{24}(0)-ie^{-\Phi(t)}\sin(Kt)\rho_S^{34}(0)\\\nonumber
+\int_0^td\tau\beta(\tau)e^{-G(\tau)}\left(\cos{K(t-\tau)}\rho_S^{13}(\tau)-i\sin{K(t-\tau)}\rho_S^{12}(\tau)\right),
\end{eqnarray}

\begin{eqnarray}
\fl \rho_S^{34}(t)
=e^{-\Phi(t)}\cos(Kt)\rho_S^{34}(0)-ie^{-\Phi(t)}\sin(Kt)\rho_S^{24}(0)\\\nonumber
+\int_0^td\tau\beta(\tau)e^{-G(\tau)}\left(\cos{K(t-\tau)}\rho_S^{12}(\tau)-i\sin{K(t-\tau)}\rho_S^{13}(\tau)\right).
\end{eqnarray}

We are going to show that the solution of the non-Markovian Master
Equation is positive. For simplicity we assume an arbitrary
X-like initial state of the two-qubit system,
\begin{eqnarray}
\fl
\rho_S(0)=p_0|00\rangle\langle00|+p_1|01\rangle\langle01|+p_2|10\rangle\langle10|+(1-p_0-p_1-p_2)|11\rangle\langle11|\\\nonumber
+C_{12}|01\rangle\langle10|+\bar{C}_{12}|10\rangle\langle01|+C_{03}|00\rangle\langle11|+\bar{C}_{03}|11\rangle\langle00|.
\end{eqnarray}
The function $G(t)$ can be re-written in the following way
\begin{equation}
G(t)=\Phi(t)+\bar{\Phi}(t)=4\sum_n |g_n|^2
\frac{\sin^2{\frac{(\epsilon-\omega_n)t}{2}}}{\left(\epsilon-\omega_n\right)^2}\geq0.
\end{equation}
After straightforward transformations we get
\begin{eqnarray}
\fl
\rho_S^{44}(t)=\left(1-\rho_S^{11}(0)-\rho_S^{22}(0)-\rho_S^{33}(0)\right)+\left(1-e^{-G(t)}\right)^2\rho_S^{11}(0)\\\nonumber+\left(1-e^{-G(t)}\right)\left(\rho_S^{22}(0)+\rho_S^{33}(0)\right),
\end{eqnarray}
taking into account the above expression for $\rho_S^{44}(t)$ and
the fact that $G(t)\geq0$ it is obvious that $\rho_S^{11}(t)$ and
$\rho_S^{44}(t)$ are nonnegative. To prove the positivity of the
solution we need to show that $\rho_S^{22}(t)$ and
$\rho_S^{33}(t)$ are nonnegative too. To this end we show that
\begin{equation}
\cos^2(Kt)\rho_S^{22}(0)+\sin^2(Kt)\rho_S^{33}(0)-\sin(2Kt)\mathrm{Im}(\rho_S^{23}(0))\geq0.
\end{equation}
Using the positivity condition for the initial density matrix
$\rho_S(0)$ which implies that $p_1p_2\geq|C_{12}|^2$ or
$\rho_S^{22}(0)\rho_S^{33}(0)\geq|\rho_S^{23}(0)|^2$ we can
strengthen the above inequality by replacing
$\sin(2Kt)\mathrm{Im}(\rho_S^{23}(0))$ by
$\pm\sin(2Kt)\sqrt{\rho_S^{22}(0)\rho_S^{33}(0)}$ and get
\begin{eqnarray}
\fl
\cos^2(Kt)\rho_S^{22}(0)+\sin^2(Kt)\rho_S^{33}(0)\pm\sin(2Kt)\sqrt{\rho_S^{22}(0)\rho_S^{33}(0)}\\\nonumber
=\left(\cos(Kt)\sqrt{\rho_S^{22}(0)}\pm\sin(Kt)\sqrt{\rho_S^{33}(0)}\right)^2\geq0.
\end{eqnarray}
Thus, from the above inequality it follows that
$\rho_S^{22}(t)\geq0$. The same statement for $\rho_S^{33}(t)$ is
established analogously. This proves that the density matrix is
positive.

\end{document}